\documentstyle[preprint,aps,eqsecnum]{revtex}

\begin{document}

\thispagestyle{empty}
\setcounter{page}{0}

\title{ CLASSICAL TRANSPORT THEORY AND HARD \\
THERMAL LOOPS IN THE QUARK-GLUON PLASMA\footnotemark[1]}

\footnotetext[1]{\baselineskip=12pt
This work is supported in part by funds provided by the
U.S.~Department of Energy (D.O.E.) under cooperative agreement
\#~DE-FC02-94ER40818. P.F.K. is supported by the Natural Sciences
and Engineering Research Council of Canada. C.L. is supported by the
Swiss National Science Foundation. C.M. is supported by the
Ministerio de Educaci\'on y Ciencia, Spain.\hfill\break
\underline{Internet addresses:}
kelly@pierre.mit.edu (P.F.K.), liu@pierre.mit.edu (Q.L.),\hfill\break
\phantom{\underline{Internet addresses:} }lucchesi@pierre.mit.edu
(C.L.), cristina@pierre.mit.edu (C.M.).\hfill\break}

\author{P.F.~Kelly,~~Q.~Liu,~~C.~Lucchesi and~~C.~Manuel}

\address{Center for Theoretical Physics,
Laboratory for Nuclear Science, and Department of Physics\\
Massachusetts Institute of Technology, Cambridge,
Massachusetts 02139}

\maketitle

\thispagestyle{empty}
\setcounter{page}{0}

\begin{abstract}
$\!\!$Classical transport theory for colored particles is investigated
and employed to derive the hard thermal loops of QCD. A formal
construction of phase-space for color degrees of freedom
is presented. The gauge invariance of the non-Abelian Vlasov
equations is verified and used as a guiding principle in our
approximation scheme. We then derive the generating functional of
hard thermal loops from a constraint satisfied at leading-order by
the color current. This derivation is more direct than alternative ones
based on perturbative quantum field theory, and shows that hard
thermal effects in hot QCD are essentially {\it classical}. As an
illustration, we analyze color polarization in the QCD plasma.
\end{abstract}

\vfill
\centerline{Submitted to: {\em Phys. Rev. D}}
\vfill

\hfill\break
\hbox to \hsize{CTP\# 2320 --- hep-ph/9406285\hfil June 1994}
\vskip-12pt
\eject

\section{INTRODUCTION}
\label{sec1}

Following a recent Letter~\cite{KLLM1}, we present an expanded and
self-contained account of the derivation of the hard thermal loops
(HTLs) of QCD from classical transport theory. In addition, we justify
the use of the {\it ad hoc} phase-space integration measure for
classical colored particles. This justification is based on the
phase-space symplectic structure, and relates directly the
(dependent) color charges to a set of (independent) Darboux
variables. We also discuss formally the gauge invariance properties
of the system of coupled non-Abelian Vlasov equations, and exploit
the gauge principle to justify the approximation scheme we use. In
order to show how physical information can be extracted, we analyze
color polarization of the quark--gluon plasma in a plane-wave {\it
Ansatz}.

We start by reviewing the work relevant to hard thermal loops in
QCD. The motivation that led to their discovery was that physical
quantities (such as damping rates) in hot QCD were
{\it gauge-dependent} when computed using the usual loop
expansion~\cite{gaugedep}. The solution to this puzzle was first
proposed by Pisarski~\cite{Pisarski1}. Subsequent development was
carried out by Braaten and Pisarski~\cite{BP1}, and by Frenkel and
Taylor~\cite{FT}. These authors realized that, in the diagrammatic
approach to high-temperature QCD, a resummation procedure is
necessary in order to take into account consistently all contributions
at leading-order in the coupling constant. Such contributions were
found to arise only from one-loop diagrams with ``soft" external  and
``hard" internal momenta. ``Soft" denotes a scale $\,\,\sim\ gT$ and
``hard" refers to one $\,\,\sim\ T$, where $g\ll 1$ is the coupling
constant, and $T$ denotes the plasma temperature. Such diagrams
were called ``hard thermal loops" in~\cite{Pisarski1,BP1}. The HTL
approach was successful in providing gauge-invariant results for
physical quantities. Identifying the momentum scales that are
relevant to the study of a hot quark--gluon plasma, as was done
in~\cite{Pisarski1,BP1}, was an essential step for all further
developments on HTLs.

An effective action for HTLs was given by Taylor and Wong
\cite{TW} who, after imposing gauge invariance, solved the resulting
condition on the generating functional. Efraty and Nair~\cite{EN}
have identified this gauge invariance condition with the equation of
motion for the topological Chern-Simons theory at zero temperature,
thereby providing a non-thermal framework for studying hard
thermal physics. Along the same line of research, the eikonal for a
Chern-Simons theory has been used by Jackiw and Nair~\cite{JN} to
obtain a non-Abelian generalization of the Kubo formula, which
governs, through the current induced by HTLs, the response of a hot
quark--gluon plasma.

Another description of hard thermal loops in QCD has been proposed
by Blaizot and Iancu~\cite{BI}. It is based on a truncation of the
Schwinger-Dyson hierarchy and yields quantum kinetic equations for
the QCD induced color current. These kinetic equations, as well as the
generating functional for HTLs, were obtained in~\cite{BI} by
performing a consistent expansion in the coupling constant, which
amounts to taking into account the coupling constant dependence
carried by the space-time derivatives. This dependence is extracted
by going to a coordinate system which separates long-wavelength,
collective excitations carrying soft momenta from the typical hard
energies of plasma particles.

Alternatively, Jackiw, Liu and Lucchesi~\cite{JLL} have shown how
HTLs can be derived from the Cornwall-Jackiw-Tomboulis composite
effective action~\cite{CJT} by requiring its stationarity, and by using
the approximation scheme developed in~\cite{BI}.

The resummation prescription of Braaten-Pisarski and
Frenkel-Taylor, as well as the consistent expansion in the coupling
constant developed by Blaizot-Iancu, although remarkably insightful,
are technically very involved and necessitate lengthy computations.
Furthermore, they are puzzling with respect to the very nature of
hard thermal loops. One wonders if a quantum field theoretical
description of hard thermal loops (involving gauge-fixing and ghost
fields) is required. Indeed, we are faced with the following situation:
in the resummation approach, HTLs emerge from loop diagrams, and
in Blaizot and Iancu's work, they arise from the Schwinger-Dyson
equations. However, hard thermal effects are UV-finite since they
are due exclusively to {\it thermal} fluctuations. One might therefore
be able to describe such effects within a {\it classical}, more
transparent, context.

This motivated us to develop a classical formalism~\cite{KLLM1} for
hard thermal loops in QCD, the natural starting point being the
classical transport theory of plasmas (see for instance~\cite {LL}). Our
effort was encouraged by the fact that for an Abelian plasma of
electrons and ions, the dielectric tensor computed~\cite{Silin} from
classical transport theory is the same as that extracted from the hard
thermal corrections to the vacuum polarization
tensor~\cite{BP1,FT,JN}. Moreover, the same situation is encountered
for non-Abelian plasmas~\cite{EH,Weldon}.

The classical transport theory for non-Abelian plasmas has been
established by Elze and Heinz~\cite {EH}, before hard thermal effects
were an issue. The HTLs of QCD were not uncovered in these early
works, mainly due to the lack of a motivation to do so and because
the transport equations had been linearized, thereby neglecting
non-Abelian contributions. There has not been, to the best of our
knowledge, any attempt to derive the complete set of HTLs for QCD
from classical transport theory. The aim of the present paper is to
give a detailed account of this derivation, the results of which have
already appeared recently in a Letter~\cite{KLLM1}.

In our approach, the generating functional of HTLs (with an arbitrary
number of soft external bosonic legs) arises as a leading-order effect
in the coupling constant. We start by reviewing the Wong equations
for classical colored particles. Following~\cite{EH}, these are
substituted into the transport equation, which governs the time
evolution of the one-particle distribution function, thereby yielding
the so-called Boltzmann equation. The latter, augmented with the
Yang-Mills equation relating the field strength to the color current,
form a consistent set of coupled, gauge-invariant differential
equations known as non-Abelian Vlasov equations.

Expanding the distribution function in powers of the coupling
constant and considering the lowest-order effects, we obtain a
constraint on the color current. The latter constraint is equivalent to
the condition found in~\cite{JLL}, and previously in~\cite{BI}, on the
induced current. The constraint on the color current leads to the
generating functional of hard thermal loops.

This work is stuctured as follows. Section~\ref{sec2} describes
classical transport theory for the quark--gluon plasma. The latter is
reviewed in Subsection~\ref{ssec2.1}. In Subsection~\ref{ssec2.2}, we
discuss and justify the {\it ad hoc} phase-space integration measure,
using Darboux variables. Subsection~\ref{ssec2.3} presents an
analysis of the gauge invariance of the system of non-Abelian Vlasov
equations. Section~\ref{sec3} contains the derivation of the hard
thermal loops of QCD. As a consequence of constraints satisfied by the
induced current (which are derived in Subsection~\ref{ssec3.1}), we
obtain the generating functional of hard thermal loops
(Subsection~\ref{ssec3.2}). In Section~\ref{sec4}, we compute the
polarization tensor from classical transport theory (at leading-order
in $g$) and extract the expression for Landau damping. The
consistency of our result with previous ones is discussed.
Section~\ref{sec5} states our conclusions. In particular, we discuss
there the validity of our approximations. In Appendix \ref{appA}, we
check the validity of the Boltzmann equation. Appendix \ref{appB}
presents a proof of the covariant conservation of the color current.

\section{CLASSICAL TRANSPORT THEORY FOR A NON-ABELIAN
PLASMA}
\label{sec2}
\subsection{Classical motion and non-Abelian Vlasov equations}
\label{ssec2.1}

The classical transport theory for the QCD plasma was developed
in~\cite{EH}, which we follow in this subsection. Consider a particle
bearing a non-Abelian $SU(N)$ color charge $Q^{a}, \ a=1,...,N^2-1$,
traversing a worldline $x^{\alpha}(\tau)$, where $\tau$ denotes the
proper time. The dynamical effects of the spin of the particles shall
be ignored, as they are typically small. The Wong equations
\cite{Wong} describe the dynamical evolution of the
variables\footnote{Note that we are using the kinetic momentum,
rather than the canonical one. A formulation in terms of canonical
variables would be equivalent~\cite{LL}.} $x^{\mu}$, $p^{\mu}$ and
$Q^{a}$:
\begin{mathletters}
\label{wongeq}
\begin{eqnarray}
m\, {{d x^{\mu}}\over{d \tau}} & = & p^{\mu}
\ , \label{wongeqa} \\[2mm]
m\, {{d p^{\mu}}\over{d \tau}} & = & g\,Q^{a}F^{\mu\nu}_{a}p_{\nu}
\ ,\label{wongeqb}\\[2mm]
m\, {{d Q^{a}}\over{d \tau}} & = & - g\,
f^{abc}p^{\mu}A^{b}_{\mu}Q^{c}\ . \label{wongeqc}
\end{eqnarray}
\end{mathletters}
$\!\!$The $f^{abc}$ are the structure constants of the group,
$F^{\mu\nu}_{a}$ denotes the field strength, $g$ is the coupling
constant, and we set $c=\hbar=k_B=1$ henceforth. Equation
(\ref{wongeqb}) is the non-Abelian generalization of the Lorentz
force law, and (\ref{wongeqc}) describes the precession in color
space of the charge in an external color field $A_\mu^a$. It is
noteworthy that the color charge $Q^a$ is itself subject to dynamical
evolution, a feature which distinguishes the non-Abelian theory from
electromagnetism.

The usual $(x,p)$ phase-space is now enlarged to $(x,p,Q)$ by
including into it color degrees of freedom for colored particles.
Physical constraints are enforced by inserting delta-functions in
the phase-space volume element $dx\,dP\,dQ$. The momentum
measure
\begin{equation}
dP = {{d^{4}p}\over{(2\pi)^{3}}}\,\,2\,\theta(p_{0})\,\,
\delta(p^{2} - m^{2})
\label{measurep}
\end{equation}
guarantees positivity of the energy and on-shell evolution. The color
charge measure enforces the conservation of the group invariants,
{\it e.g.}, for $SU(3)$,
\begin{equation}
dQ = d^8 Q\,\, \delta(Q_{a}Q^{a} - q_{2})\,\,
\delta(d_{abc}Q^{a}Q^{b}Q^{c} - q_{3}) \ ,
\label{measureq}
\end{equation}
where the constants $q_{2}$ and $q_{3}$ fix the values of the
Casimirs and $d_{abc}$ are the totally symmetric group constants.
The color charges which now span the phase-space are dependent
variables. These can be formally related to a set of independent
phase-space Darboux variables. This derivation is presented in
Subsection \ref{ssec2.2} below.

The one-particle distribution function $f(x,p,Q)$ denotes the
probability for finding the particle in the state $(x,p,Q)$. It evolves in
time via a transport equation,
\begin{equation}
m\, {{d f(x,p,Q)}\over{d \tau}} = C[f](x,p,Q) \ , \label{transport}
\end{equation}
where $C[f](x,p,Q)$ denotes the collision integral, which we
henceforth set to zero. Using the equations of motion~(\ref{wongeq}),
(\ref{transport}) becomes, in the collisionless case, the Boltzmann
equation:
\begin{equation}
p^{\mu}\left[{{\partial}\over{\partial x^{\mu}}}
- g\, Q_{a}F^{a}_{\mu\nu}{{\partial}\over{\partial p_{\nu}}}
- g\, f_{abc}A^{b}_{\mu} Q^{c}{{\partial}\over{\partial Q_{a}}}
\right] f(x,p,Q) = 0 \ . \label{boltzmann}
\end{equation}
In Appendix \ref{appA}, an explicit microscopic distribution function
is presented and used to check the validity of (\ref{boltzmann}).

A complete, self-consistent set of non-Abelian Vlasov equations for
the distribution function and the mean color field is obtained by
augmenting the Boltzmann equation with the Yang-Mills equations:
\begin{equation}
[D_\nu F^{\nu\mu}]^a(x) = J^{\mu\, a}(x)\ .
\label{yangmills}
\end{equation}
The covariant derivative is defined as
$D_\mu^{ac} =\partial_\mu \delta^{ac} + g\, f^{abc} A_\mu^b$.
The total color current $J^{\mu\,a}(x)$ is given by the sum of all
contributions from particle species and helicities,
\begin{equation}
J^{\mu\, a}(x) = \sum_{\rm species}\ \sum_{\rm helicities}\
j^{\mu\, a}(x)\ .
\label{sumsum}
\end{equation}
Each $j^{\mu\,a}(x)$ (species and spin indices are implicit) is
computed from the corresponding distribution function as
\begin{equation}
j^{\mu\,a} (x) = g\, \int dPdQ\ p^\mu Q^a f(x,p,Q)
\label{cr5}
\end{equation}
and it is covariantly conserved,
\begin{equation}
\left(D_{\mu} j^{\mu} \right)^a (x) = 0\ ,
\label{cr6}
\end{equation}
as can be checked by using the Boltzmann equation (a detailed proof
is presented in Appendix \ref{appB}). For later convenience, we
define the total and individual current momentum-densities:
\begin{equation}
J^{\mu\, a}(x,p) = \sum_{\rm species}\ \sum_{\rm helicities} \
j^{\mu\, a}(x,p)\ ,\qquad
j^{\mu\,a} (x,p) =  g\,\int dQ\ p^\mu Q^a f(x,p,Q)\ .
\label{cudens}
\end{equation}

Note that a solution to the set of Vlasov equations
(\ref{boltzmann})--(\ref{yangmills}) is specified by giving the
forms for the gauge potential $A_\mu(x)$ and for the distribution
function $f(x,p,Q)$.

\subsection{Phase-space for colored particles}
\label{ssec2.2}

In order to carry out the transport theory analysis for classical
colored particles, it has been necessary to extend phase-space by the
addition of the color charges. In (\ref{measureq}), the charges are
constrained to remain within the group manifold by means of
delta-functions which fix the values of the
(representation-dependent) group Casimirs. In fact, at an operational
level, this is the approach adopted in the rest of the present paper.
In this subsection, we formally justify this approach by analysis of
the symplectic structure of the group manifold~\cite{AFS,Johnson}.
We work out explicitly the $SU(2)$  and $SU(3)$ cases.

The group $SU(2)$, is generated by three charges,
$(Q_{1},Q_{2},Q_{3})$, and has one Casimir,  $Q^{a}Q_{a}$. The
structure constants are $f_{abc} = \epsilon_{abc}$, while $d_{abc} =
0$. From the point of view adopted throughout the rest of this paper
the phase-space color measure is
\begin{equation}
dQ =  dQ_{1}\, dQ_{2}\, dQ_{3}\,\delta(Q^{a}Q_{a} - q_{2}) \ ,
\label{meas-su2}
\end{equation}
where $q_{2}$ denotes the value of the quadratic Casimir.

New coordinates $(\phi,\pi,J)$ may be introduced by the following
transformation~\cite{Johnson}:
\begin{equation}
Q_{1} = \cos\phi \ \sqrt{J^{2} - \pi^{2}} \ , \qquad
Q_{2} = \sin\phi \ \sqrt{J^{2} - \pi^{2}} \ , \qquad
Q_{3} = \pi \ .
\label{gens-su2}
\end{equation}
Note that $\pi$ is bounded, $ -J \leq \pi \leq J$. That the group
manifold has spherical geometry is readily apparent if one chooses
$\pi = J\cos\theta$. The variables $\phi$ and $\pi$ form a
canonically conjugate pair; the Poisson bracket may be formed in the
conventional manner:
\begin{equation}
\{A,B\}_{\rm PB} \equiv {{\partial A}\over{\partial \phi}}
{{\partial B}\over{\partial \pi}} -
{{\partial A}\over{\partial \pi}}
{{\partial B}\over{\partial \phi}} \ . \label{pb}
\end{equation}
It is easily verified that the charges as given by (\ref{gens-su2})
form a representation of $SU(2)$ under the Poisson bracket, {\it i.e.},
\begin{equation}
\{Q_{a},Q_{b}\}_{\rm PB} = \epsilon_{abc}\ Q_{c}\ .
\label{pb-su2}
\end{equation}

The above Poisson bracket structure allows one to identify $\phi$
and $\pi$ as Darboux variables (see for instance~\cite{AFS}). The
Jacobian of the transformation from $(Q_{1},Q_{2},Q_{3})$ to
$(\phi,\pi,J)$ takes the value
\begin{equation}
\left|{{\partial (Q_{1},Q_{2},Q_{3})} \over {\partial (\phi,\pi,J)}}\right|
= J \ . \label{jac-su2}
\end{equation}
Performing the change of variables (\ref{gens-su2}) in
(\ref{meas-su2}) and substituting the value of the quadratic Casimir,
$Q^{a}Q_{a} = J^2$, the color measure reads
\begin{equation}
dQ = d\phi\, d\pi \, dJ \, J \, \delta(J^{2} - q_{2})
\ , \label{measD-su2}
\end{equation}
which, upon integration over the constrained variable $J$, is just the
proper, canonical volume element $d\phi\, d\pi$, up to an irrelevant
constant.

The group $SU(3)$ has eight charges, $(Q_{1}, \ldots , Q_{8})$ and two
conserved quantities, the quadratic and the cubic Casimirs,
$Q^{a}Q_{a}$ and $d_{abc}Q^{a}Q^{b}Q^{c}$, respectively. The
phase-space color measure is quoted above in (\ref{measureq}).

As in the $SU(2)$ case, new coordinates
$(\phi_{1},\phi_{2},\phi_{3},\pi_{1},\pi_{2},\pi_{3},J_1,J_2)$ may be
introduced by means of the following transformations~\cite{Johnson}:
\begin{equation}
\begin{array}{rclcrcl}
Q_{1} &=& \cos\phi_{1}\, \pi_{+}\,\pi_{-} \ , &&
Q_{2} &=& \sin\phi_{1}\, \pi_{+}\,\pi_{-} \ , \nonumber\\
Q_{3} &=& \pi_{1} \ , &&&&\nonumber\\
Q_{4} &=& C_{++}\,\pi_{+}\,A + C_{+-}\,\pi_{-}\,B \ , &\qquad&
Q_{5} &=& S_{++}\,\pi_{+}\,A + S_{+-}\,\pi_{-}\,B \ , \label{gens-
su3}\\
Q_{6} &=& C_{-+}\,\pi_{-}\,A - C_{--}\,\pi_{+}\,B \ , &\qquad&
Q_{7} &=& S_{-+}\,\pi_{-}\,A - S_{--}\,\pi_{+}\,B \ , \nonumber\\
Q_{8} &=& \pi_{2} \ , &&&&\nonumber
\end{array}
\end{equation}
in which we have used the definitions:
\begin{equation}
\begin{array}{rclcrcl}
\pi_+ &=& \sqrt{\pi_3 +\pi_1}\ ,&\qquad&
\pi_- &=& \sqrt{\pi_3 -\pi_1}\ , \nonumber\\[2mm]
C_{\pm\pm} &=& \cos\left[\frac{1}{2}(\pm\phi_1
+\sqrt{3}\phi_{2}\pm\phi_{3})\right]\ , &\qquad&
S_{\pm\pm} &=& \sin\left[\frac{1}{2}(\pm\phi_1
+\sqrt{3}\phi_2\pm\phi_3)\right] \ , \nonumber
\end{array}
\end{equation}
and $A,\ B$ are given by
\begin{eqnarray}
A &=& \frac{1}{2\pi_{3}} \sqrt{
\left(\frac{J_1 - J_2}{3} + \pi_{3} + \frac{\pi_{2}}{\sqrt{3}}\right)
\left(\frac{J_1 + 2J_2}{3} + \pi_{3} + \frac{\pi_{2}}{\sqrt{3}}\right)
\left(\frac{2J_1 + J_2}{3} - \pi_{3} - \frac{\pi_{2}}{\sqrt{3}}\right) }
\quad , \nonumber\\[3mm]
B &=& \frac{1}{2\pi_{3}} \sqrt{
\left(\frac{J_2 - J_1}{3} + \pi_{3} - \frac{\pi_{2}}{\sqrt{3}}\right)
\left(\frac{J_1 + 2J_2}{3} - \pi_{3} + \frac{\pi_{2}}{\sqrt{3}}\right)
\left(\frac{2J_1 + J_2}{3} + \pi_{3} - \frac{\pi_{2}}{\sqrt{3}}\right) }
\quad .\nonumber
\end{eqnarray}
Note that in this representation, the set $(Q_{1},Q_{2},Q_{3})$ forms
an $SU(2)$ subgroup with quadratic Casimir
$Q_{1}^{2} + Q_{2}^{2} + Q_{3}^{2} = \pi_{3}^{2}$.
It can be verified that the expressions above for $Q_{1}, \ldots,
Q_{8}$ form a representation of the group $SU(3)$:
\begin{equation}
\{Q_{a},Q_{b}\}_{\rm PB} = f_{abc}Q_{c} \ , \label{pb-su3}
\end{equation}
under the Poisson bracket
\begin{equation}
\{A,B\}_{\rm PB} \equiv \sum_{i = 1}^{3} \left(
{{\partial A} \over {\partial\phi_{i}}}
{{\partial B} \over {\partial \pi_{i}}} -
{{\partial A} \over {\partial \pi_{i}}}
{{\partial B} \over {\partial \phi_{i}}} \right)
\ , \label{pb-gen}
\end{equation}
where the canonical pairs are $\{ \phi_i, \pi_i \}_{i = 1, 2, 3}$.

As is implicit in the above, the two Casimirs depend only on $J_1$
and $J_2$. They can be computed, using the values given in the table
below, as:
\begin{mathletters}
\label{cas-su3}
\begin{eqnarray}
Q^{a}Q_{a} &=& \frac{1}{3} (J_1^2+J_1J_2+J_2^2) \ ,
\label{c2-su3} \\[3mm]
d_{abc}Q^{a}Q^{b}Q^{c} &=& \frac{1}{18}(J_1-J_2)
(J_1+2J_2)(2J_1+J_2) \ .
\label{c3-su3}
\end{eqnarray}
\end{mathletters}

\vspace{5mm}
\begin{tabular}{c|cccccccccccccccc}\hline\hline
$d_{abc}$
&$d_{118}$
&$d_{146}$
&$d_{157}$
&$d_{228}$
&$d_{247}$
&$d_{256}$
&$d_{338}$
&$d_{344}$
&$d_{355}$
&$d_{366}$
&$d_{377}$
&$d_{448}$
&$d_{558}$
&$d_{668}$
&$d_{778}$
&$d_{888}$
\\ \hline
${\rm Value}$
&$1\over\sqrt{3}$
&$1\over 2$
&$1\over 2$
&$1\over\sqrt{3}$
&$-{1\over 2}$
&$1\over 2$
&$1\over\sqrt{3}$
&$1\over 2$
&$1\over 2$
&$-{1\over 2}$
&$-{1\over 2}$
&$-{1\over 2\sqrt{3}}$
&$-{1\over 2\sqrt{3}}$
&$-{1\over 2\sqrt{3}}$
&$-{1\over 2\sqrt{3}}$
&$-{1\over\sqrt{3}}$
\\ \hline\hline
\end{tabular}
\vspace{3mm}

\hspace{20mm}{\bf Table:} Values of the (non-zero) $SU(3)$
totally symmetric constants.
\vspace{4mm}

The phase-space color measure for $SU(3)$, given in
(\ref{measureq}), may be transformed to the new coordinates
through use of (\ref{cas-su3}) and evaluation of the Jacobian
\begin{equation}
\left|\frac{\partial (Q_{1},Q_{2},\ldots,Q_{8})}
{\partial
(\phi_{1},\phi_{2},\phi_{3},\pi_{1},\pi_{2},\pi_{3},J_1,J_2)}\right|
= \frac{\sqrt{3}}{48}\ J_1\,J_2\,(J_1+J_2) \ .\label{jac-su3}
\end{equation}
The measure reads:
\begin{eqnarray}
dQ = d\phi_{1}\,d\phi_{2}\,d\phi_{3}\,d\pi_{1}\,d\pi_{2}
\,d\pi_{3}\,dJ_1\,dJ_2\ &&\frac{\sqrt{3}}{48}\ J_1\,J_2\,(J_1+J_2) \
\delta\Bigl(\frac{1}{3} (J_1^2+J_1J_2+J_2^2)- q_{2}\Bigr)\
\times\nonumber\\
&&\delta\Bigl(\frac{1}{18}(J_1-J_2)(J_1+2J_2)(2J_1+J_2) - q_{3}
\Bigr)\ . \label{foo2}
\end{eqnarray}
Since the two Casimirs are linearly independent, the delta-functions
uniquely fix both $J_1$ and $J_2$ to be representation-dependent
constants. Upon integrating over $J_1$ and $J_2$, (\ref{foo2})
reduces to a constant times the proper canonical volume element
$\prod_{i=1}^3\,d\phi_i\,d\pi_i$.

The construction of the canonical phase-space measure for the
general $SU(N)$ case is a departure from our purposes and will not
be undertaken here. Nevertheless, based on the examples we treated
explicitly, it is apparent that no difficulties will arise for
$N>3$~\cite{AFS}.

In principle, the classical transport theory analysis should be carried
out using canonical, independent integration variables and the
phase-space volume element should be taken to be the proper
canonical volume element. In this subsection, we have shown the
equivalence of the {\it ad hoc} phase-space color measure and the
proper canonical volume element. Hence, the use of the color charges
as phase-space coordinates is justified.

\subsection{Gauge invariance of the non-Abelian Vlasov equations}
\label{ssec2.3}

Before addressing the question of the gauge invariance of the system
of Vlasov equations, we consider the Wong equations (\ref{wongeq}).
These are invariant under the finite gauge
transformations\footnote{We use here matrix notation, {\it e.g.}
$Q=Q_a t^a$, ${\partial\over \partial Q}={\partial\over\partial
Q^a}\,t^a$, where the generators are represented by antihermitian
matrices $t^a$ in the fundamental representation,
$[t^a,t^b]=f^{abc}t^c$, and they are normalized as $~{\rm Tr}~ (t^a t^b)=
- {1 \over 2} \,\delta^{ab}$.}:
\begin{mathletters}
\label{gaugetrsf}
\begin{eqnarray}
\bar{x}^{\mu}&=&x^{\mu}\ ,\\
\bar{p}^{\mu}&=&p^{\mu}\ , \\
\bar{Q}&=& U \,Q \,U^{-1}\ , \\
{\bar A}_\mu &=& U\,A_\mu \,U^{-1}-{1\over g}\,U\,
{\partial\over \partial x_\mu}\,U^{-1}\ ,
\end{eqnarray}
\label{gtrsf}
\end{mathletters}
$\!\!$where $U(x)={\rm exp}[-g\,\varepsilon^a(x)\,t^a]$ is a group
element.

Accordingly, the derivatives appearing in the Boltzmann equation
(\ref{boltzmann}) transform as:
\begin{mathletters}
\begin{eqnarray}
{\partial\over\partial x^\mu}&=&
{\partial\over\partial\bar{x}^\mu}
- 2 ~{\rm Tr}~ \Biggl([\ ({\partial\over\partial {\bar x}^\mu}U)
U^{-1}\ ,\  \bar{Q}\ ]
{\partial\over\partial\bar{Q}}\Biggr)\label{eq:gauge2a}\\
{\partial\over\partial p^\mu}&=&
{\partial\over\partial\bar{p}^\mu}\label{eq:gauge2b}\\
{\partial\over\partial Q}&=&
U^{-1} {\partial\over\partial\bar{Q}}U\label{eq:gauge2c}\ .
\end{eqnarray}
\label{eq:gauge2}
\end{mathletters}
$\!\!$Consequently, the Boltzmann equation (rewritten here in terms
of traces):
\begin{equation}
\left[p^\mu {\partial\over\partial x^\mu}- 2\,g\,
p^\mu ~{\rm Tr}~(Q F_{\mu\nu})
{\partial\over\partial p_\nu}+ 2\,g\,p^\mu
{}~{\rm Tr}~\Bigl([\ A_\mu\ ,\ Q\ ]
{\partial\over\partial Q}\Bigr)\right]\ f(x, p, Q)=0
\label{eq:gauge3}
\end{equation}
becomes, in the new coordinates:
\begin{eqnarray}
\Biggl[
&&\bar{p}^{\mu} {\partial \over \partial \bar{x}^{\mu}}
- 2\,g\,\bar{p}^{\mu}~{\rm Tr}~\Bigl(
[\ ({\partial \over \partial {\bar x}^{\mu}}U) U^{-1}\ ,\ \bar{Q}\ ]
{\partial \over \partial \bar{Q}}\Bigr)
+ 2 \,g \,\bar{p}^{\mu} ~{\rm Tr}~(\bar{Q} \bar{F}_{\mu \nu})
{\partial \over \partial \bar{p}_{\nu}}
\nonumber\\
&&+ 2\,g\,\bar{p}^{\mu} ~{\rm Tr}~\Bigl(
[\ \bar{A}_\mu+{1\over g}({\partial\over\partial{\bar x}^\mu}U)U^{-
1}\ ,\ \bar{Q}\ ] {\partial \over \partial \bar{Q}}\Bigr)\Biggr]
\ \bar{f}(\bar{x}, \bar{p}, \bar{Q})=0\ ,
\label{eq:gauge4}
\end{eqnarray}
where we have defined
\begin{equation}
\bar{f}(\bar{x}, \bar{p}, \bar{Q})=
f\Bigl(x(\bar{x}, \bar{p}, \bar{Q}), p(\bar{x}, \bar{p}, \bar{Q}),
Q(\bar{x}, \bar{p}, \bar{Q})\Bigr)\ .
\label{fbar}
\end{equation}
Simplifying (\ref{eq:gauge4}), we obtain:
\begin{equation}
\left[\bar{p}^{\mu} {\partial \over \partial \bar{x}^{\mu}}
+ 2\,g\,\bar{p}^{\mu} ~{\rm Tr}~(\bar{Q} \bar{F}_{\mu \nu})
{\partial \over \partial \bar{p}_{\nu}}
+ 2\,g\,\bar{p}^{\mu} ~{\rm Tr}~\left([\ \bar{A}_{\mu}\ , \ \bar{Q}\ ]
{\partial \over \partial \bar{Q}}\right)\right]\ \bar{f}(\bar{x},
\bar{p},
\bar{Q})=0\ .\label{eq:gauge5}
\end{equation}
This proves that the Boltzmann equation is invariant under gauge
transformations. On the other hand, the Yang-Mills equation
(\ref{yangmills}) is gauge-covariant. Indeed, the color current
(\ref{cr5}) transforms under (\ref{gtrsf}) as a gauge covariant vector:
$j^{\mu}(x)\rightarrow{\bar j}^{\mu}({\bar x})=\int d{\bar
P}\,d{\bar Q}\,{\bar p}^\mu \,{\bar Q} \,\bar{f}(\bar{x}, \bar{p},
\bar{Q})$. Due to the gauge-invariance of the phase-space measure,
to the transformation property of $f$ (\ref{fbar}), and to
(\ref{gtrsf}), ${\bar j}^{\mu}(x)$ may be rewritten as:
\begin{equation}
{\bar j}^{\mu}({\bar x})=\int dP\,dQ\,p^\mu\,U\,Q\,
U^{-1}\,f(x,p,Q)=U\,j^{\mu}(x)\,U^{-1}\ .
\end{equation}
Hence, the system of non-Abelian Vlasov equations is
gauge-covariant, with the distribution function $f(x,p,Q)$
transforming as a scalar. Note that the gauge symmetry also implies
that the gauge transform $\{ {\bar A}_\mu(x),{\bar f}(x,p,Q)\}$ of a
set of solutions $\{ A_\mu(x), f(x,p,Q)\}$ to the Vlasov equations:
\begin{mathletters}
\begin{eqnarray}
&&{\bar A}_\mu(x) = U\,A_\mu(x)\,U^{-1} -{1\over
g}\,U\,{\partial\over\partial x^\mu}\,U^{-1}\ ,\\
&&{\bar f}(x,p,Q) = f (x,p,U\,Q\,U^{-1})
\label{trsfsol}
\end{eqnarray}
\end{mathletters}
$\!\!$is still a solution.

\section{EMERGENCE OF HARD THERMAL LOOPS}
\label{sec3}

\subsection{Constraint on the color current}
\label{ssec3.1}

Classical transport theory is now employed to study soft excitations
in a hot, color-neutral quark--gluon plasma. In the high-temperature
limit, the masses of the particles can be neglected and shall
henceforth be assumed to vanish. The wavelength of a soft excitation
is of order ${1\over g\,|A|}$ and the coupling constant $g$ is assumed
to be small. We then expand the distribution function $f(x,p,Q)$ in
powers of $g$:
\begin{equation}
f=f^{(0)}+gf^{(1)}+g^2f^{(2)}+...\ ,
\label{L1}
\end{equation}
where $f^{(0)}$ is the equilibrium distribution function in the
absence of a net color field, and is given by:
\begin{equation}
f^{(0)}(p_0)=C\ n_{B,F}(p_0)\ .
\end{equation}
Here $C$ is a normalization constant and $n_{B,F}(p_0)=1/(e^{\beta
|p_0|}\mp 1)$ is the bosonic, resp. fermionic, probability distribution.

At leading-order in $g$, the color current (\ref{cudens}) is
\begin{equation}
j^{\mu\,a}(x,p)=g^2 \int dQ\  p^{\mu} Q^a f^{(1)}(x,p,Q)\ ,
\label{L2}
\end{equation}
while the Boltzmann equation (\ref{boltzmann}) reduces to
\begin{equation}
p^{\mu} \left({\partial\over\partial x^{\mu}}-g\, f^{abc} A_{\mu}^b
Q_c {\partial\over\partial Q^a}\right)
f^{(1)}(x,p,Q) = p^{\mu} Q_a F_{\mu \nu}^a {\partial\over \partial
p_{\nu}} f^{(0)}(p_0)\ .
\label{L3}
\end{equation}
Due to the softness of the excitation, the ${\partial\over\partial
x^{\mu}}$ in the above equation is of order $g\,|A|$, so we are taking
into account consistently all contributions of order $g$. The
approximation we use guarantees that the non-Abelian gauge
symmetry of the exact Boltzmann equation (\ref{boltzmann}) is
preserved in the approximate equation (\ref{L3}). As a consequence
$f^{(0)}$ and $f^{(1)}$, like $f$, transform separately as
gauge-invariant scalars. Other approximations, which have been
carried out in the past~\cite{EH}, have discarded the non-Abelian
contributions, thereby breaking the non-Abelian gauge symmetry of
the Boltzmann equation.

The equations (\ref{L2}) and (\ref{L3}) yield the following constraint
on the color current:
\begin{equation}
[\,p \cdot D\,\, j^{\mu}(x,p)]^a = g^2\, p^{\mu} p^{\nu} F_{\nu \rho}^b
{\partial\over\partial p_{\rho}}
\left(\int dQ\ Q^a Q_b f^{(0)}(p_0)\right) \ ,
\label{L4}
\end{equation}
where, from color symmetry, we have$\int dQ\ Q^a Q_b
f^{(0)}(p_0)=C_{B,F}\,\, n_{B,F}(p_0)\, \delta^a_{\,b}$ with $C_{B}=N,\
C_F={1\over 2}$ for gluons, resp. fermions. Thus, upon summation
over all species ($N_F$ quarks, $N_F$ antiquarks and one
[$(N^2-1)$-plet] gluon) and helicities (2 for quarks-antiquarks and
for the massless gluon), (\ref{L4}) yields,
\begin{equation}
[\,p \cdot D\,\, J^{\mu}(x,p)]^a = 2\,g^2\, p^{\mu} p^{\nu}
F_{\nu 0}^a {d \over dp_0}[N\, n_B(p_0)+N_F\, n_F(p_0)] \ .
\label{L5}
\end{equation}
Similar results have been obtained in~\cite{BI,JLL}, in a quantum
field theoretic setting.

\subsection{Derivation of hard thermal loops}
\label{ssec3.2}

Subsequent steps which lead to the generating functional of HTLs
have been described in~\cite{JLL}, the results of which were used
straightforwardly in~\cite{KLLM1}, for the sake of brevity. Here, we
present a simpler derivation of HTLs by exploiting fully the structure
of the momentum integration measure (\ref{measurep}).

We first integrate equation (\ref{L5}) over $|{\bf p}|$ and $p_0$
using the massless limit of the momentum measure $dP$
(\ref{measurep}). Therefore, the (massless) mass-shell constraint
enforces ${\bf |p|}=p_0$, and we thus introduce the unit vector
${\bf\hat p}\equiv {\bf p/|p|}$. Introducing also $v\equiv
(1,{\bf\hat p})$, the integration of (\ref{L5}) yields (group indices
are henceforth omitted):
\begin{equation}
v\cdot D \ {\cal J}^{\mu}(x,v)=- 2  \, \pi^2
 \,m^2_D\, v^\mu\, v^\rho\, F_{\rho 0}(x)\ ,
\label{3.103}
\end{equation}
where $m_D$ is the Debye screening mass
\begin{equation}
m_D=gT\,\sqrt{{N+N_F/2\over 3}}\ ,
\label{debye}
\end{equation}
and we have defined
\begin{equation}
{\cal J}^\mu (x,v) = \int|{\bf p}|^2\, d|{\bf p}|\,dp_0\,2\,\theta (p_0)\,
\delta(p^2)\,\,J^\mu(x,p)\ .
\label{3.1031}
\end{equation}

Notice (for later use) that, using $\int dP = \int {d\Omega\over
(2\pi)^3} |{\bf p}|^2 d|{\bf p}|dp_0\, 2\theta(p_0)\,\delta(p^2)$,
where $d\Omega$ denotes integration over all angular directions of
the unit vector ${\bf\hat p}$, we can rewrite the expression
$J^\mu(x)=\int dP J^\mu(x,p)$ for the color current as
\begin{equation}
J^\mu (x)=\int{d\Omega\over (2\pi)^3}\, {\cal J}^\mu(x,v)\ .
\label{3.1041}
\end{equation}
After decomposing ${\cal J}^{\mu}(x,v)$ as
\begin{equation}
{\cal J}^{\mu}(x,v)={\tilde {\cal J}}^{\mu}(x,v) - 2
\,\pi^2 \,m^2_D\, v^\mu A_0(x)\ ,
\label{3.13}
\end{equation}
we get as our final condition on the color current:
\begin{equation}
v\cdot D \  {\tilde {\cal J}}^{\mu}(x,v)=2\, \pi^2 \ m^2_D \ v^\mu
{\partial\over\partial x^0} \,\Bigl(v\cdot A(x)\Bigr)\ .
\label{3.14}
\end{equation}

It has been shown that solutions to (\ref{3.14}) can be obtained from
a functional $W(A,v)$ as~\cite{TW}
\begin{equation}
{\tilde {\cal J}}^{\mu}(x,v)={{\delta W(A,v)}\over
{\delta A_\mu(x)}}\ .
\label{3.15}
\end{equation}
Equation (\ref{3.14}) then implies that $W(A,v)$ depends only on $\
A_+\equiv v\cdot A$, {\it i.e.} $W(A,v) =W(A_+)$, and ${\tilde {\cal
J}}^{\mu}={{\delta W( A_+ )}\over{\delta A_+ }}\,v^\mu$. In turn,
$W( A_+ )$ satisfies, as a consequence of (\ref{3.14}),
\begin{equation}
v \cdot D \ {{\delta W( A_+ )}\over{\delta  A_+ }}= 2\, \pi^2 \
m^2_D \ {\partial\over\partial x^0}\, A_+  \ .
\label{3.16}
\end{equation}
By introducing new coordinates $(x_+,x_-,{\bf x}_\bot )$,
\begin{equation}
x_+= {\bar v}\cdot x\ ,\qquad  x_-=v \cdot x\ ,\qquad
{\bf x}_\bot ={\bf x}-({\bf\hat p}\cdot{\bf x}){\bf\hat p}\ ,
\label{3.17}
\end{equation}
with $\bar{v} \equiv (1,-{\bf\hat{p}})$ and ${\bf x}_\bot \cdot
{\bf\hat p} =0$, we can rewrite $v \cdot {\partial\over\partial x}$
as $\partial_+$ and (\ref{3.16}) becomes:
\begin{equation}
\partial_+\,{\delta W(A_+)\over \delta A_+}
+g\,\left[ A_+,{\delta W(A_+)\over \delta A_+}\right]
=2\,\pi^2\,m_D^2\, {\partial\over\partial x^0} A_+\ .
\label{3.26}
\end{equation}

Now using (\ref{3.1041}), (\ref{3.13}) and (\ref{3.15}), we define an
effective action $\Gamma$ that generates the color current, {\it i.e.},
$J^\mu(x)=-\frac{\delta\Gamma[A(x)]}{\delta A_\mu(x)}$, where
$\Gamma$ takes the form:
\begin{equation}
\Gamma ={m_D^2\over 2}\,\int d^4\!x\,A^a_0(x)A^a_0(x)
-\int {d\Omega\over (2\pi)^3} \,W(A_+)\ .
\label{Gamma}
\end{equation}
This is the expression for the effective action generating hard
thermal loops~\cite{BP1,FT}, while equation (\ref{3.26}) represents
the condition of gauge invariance~\cite{TW} for this generating
functional. By solving (\ref{3.26}), Taylor and Wong~\cite{TW}, as
well as Efraty and Nair~\cite{EN}, have given an explicit form for the
functional $W(A_+)$ in the second term of (\ref{Gamma}). The first
term is a mass term for $A_0^a(x)$ and describes Debye screening.

This concludes our derivation of the hard thermal loops of QCD from
classical transport theory.

\section{APPLICATION: COLOR POLARIZATION}
\label{sec4}

As an application of the classical transport formalism presented
above, we solve the approximate Boltzmann equation (\ref{L3}) for
plane-wave excitations in a collisionless isotropic plasma of quarks
and gluons. Recall that in the case of  a collisionless plasma of
electrons and ions the Abelian version of equation (\ref{L3}) has
been solved exactly for an electromagnetic
plane-wave~\cite{LL,Silin}, making it possible to  study the response
of an Abelian plasma to a weak field. We shall proceed analogously
in the non-Abelian case. We consider a plane-wave {\it Ansatz} in
which the vector gauge fields  only depend  on $x^\mu$ through the
combination $x \cdot k$, where $k^\mu = (\omega, {\bf k})$ is the
wave vector, {\it i.e.}, $A_{\mu}^{a} (x) \equiv A_{\mu}^{a} (k \cdot
x)$. With this {\it Ansatz} (which has been used in \cite{BI2} to
study the non-Abelian Kubo equation) the solution of (\ref{L3}) is
\begin{equation}
f^{(1)} (x,p,Q) = Q_a \left (  A_0 ^a (x) - \omega \,
\frac { p \cdot A^a (x) } {p \cdot  k} \right) \frac {d}{dp_0} f^{(0)}
(p_0) \ .
\end{equation}
Hence, the color current is given by
\begin{equation}
j^{\mu} _a (x) = g^2 \int dP\, dQ \,p^{\mu} Q_a Q_b
\left ( A_0 ^b (x) -\omega \, \frac { p \cdot A^b (x)}{p \cdot k}\right)
\frac{d}{dp_0}f^{(0)} (p_0) \ .
\end{equation}
The integration over color charges can be done  by using $\int dQ\
Q^a Q_b =C_{B,F}\, \delta^a_{\,b}$ with $C_{B}=N,\ C_F={1\over 2}$ for
gluons, resp. fermions. The integration over $p_0$ and $|{\bf p}|$ is
straighforward as well. Upon summation over all species and
helicities (see Section III for notations and conventions), we get the
following expression for the total color current:
\begin{equation}
J^{\mu}_a (x) =  m_D ^2 \int \frac {d \Omega} {4 \pi} \, v^{\mu}
\left (\omega \, \frac{ v \cdot A_a (x)} {v \cdot k}- A^0 _a (x) \right)
\ .
\label{curr}
\end{equation}

The  polarization tensor $\Pi^{\mu\nu}_{a b}$ can be computed from
(\ref{curr}) by using the relation
\begin{equation}
J^{\mu}_a (x) = \int d^4 y \, \Pi ^{\mu\nu} _{a b} (x-y)\,
A_{\nu} ^{b} (y) \ .
\end{equation}
It reads:
\begin{equation}
\Pi^{\mu\nu}_{ab} (x-y)=  m_D ^2 \left (-g^{\mu 0} g^{\nu 0} +
\omega \, I^{\mu\nu} (\omega,{\bf k}) \right) \,\delta^{(4)} (x-y) \
\delta_{ab} \ ,
\label{polar}
\end{equation}
where $I^{\mu\nu}$ is defined as
\begin{equation}
I^{\mu\nu} (\omega,{\bf k}) = \int \frac{d \Omega}{4 \pi} \,
\frac {v^{\mu} v^{\nu}} {\omega - \bf {k} \cdot \bf{v}} \,\ .
\end{equation}
To avoid the poles in the above  integrand, we impose retarded
boundary conditions, {\it i.e.}, we replace $\omega$ by  $\omega + i
\epsilon$.  Using the identity
\begin{equation}
\frac{1}{z+i \epsilon} = {\cal{P}} \,\frac{1}{z} - i\pi\, \delta(z) \ ,
\end{equation}
where $\cal{P}$ stands for the principal value, the real and
imaginary parts of the polarization tensor are
\begin{mathletters}
\begin{eqnarray}
{\rm Re} \, \Pi^{\mu\nu}_{ab} (\omega,{\bf k}) & = & - \delta_{ab}\
m_D^2 \left (-g^{\mu 0} g^{\nu 0} + \omega \, {\cal{P}}  \int \frac{d
\Omega}{4 \pi} \, \frac {v^{\mu} \,v^{\nu}} {\omega  - \bf {k} \cdot
\bf{v}} \right) \ , \label{realpo}\\
{\rm Im} \, \Pi^{\mu\nu}_{ab} (\omega,{\bf k}) & = & - \delta_{ab}\
m_D^2 \,\pi\,\omega\int \frac{d \Omega}{4 \pi}\, v^{\mu}\, v^{\nu}\,
\delta(\omega - \bf{k} \cdot \bf{v}) \ .
\label{landaudamp}
\end{eqnarray}
\end{mathletters}

The imaginary part of the polarization tensor (\ref{landaudamp})
describes Landau damping in the quark--gluon plasma. Explicitly:
\begin{mathletters}
\begin{eqnarray}
{\rm Im} \, \Pi^{00}_{ab} (\omega,{\bf k}) & = & - \delta_{ab} \,
m^2 _D \, \pi\ \frac{\omega}{2 |{\bf k}|}\, \theta(|{\bf k}|^2 -
\omega^2) \ , \\
{\rm Im} \, \Pi^{0i}_{ab} (\omega,{\bf k}) & = &- \delta_{ab} \,
m^2 _D \, \pi\ \frac{\omega^2}{2 |{\bf k}|^2} \frac{k^i}{|{\bf k}|} \,
\theta (|{\bf k}|^2 - \omega^2) \ , \\
{\rm Im} \, \Pi^{ij}_{ab} (\omega,{\bf k}) & = & - \delta_{ab} \,
m^2 _D \, \pi\, \left[\frac{\omega^2}{4 |{\bf k}|^2} \!\!\left(
\frac{|{\bf k}|}{\omega} - \frac{\omega}{|{\bf k}|} \right)\!\!\left(
\delta^{ij} - \frac{k^i k^j}{|{\bf k}|^2} \right)\! + \frac{\omega^3}{2
|{\bf k}|^3}\frac{k^i k^j}{|{\bf k}|^2} \right]\!\theta (|{\bf k}|^2\! -
\omega^2) \ .
\end{eqnarray}
\end{mathletters}
$\!\!$From the above $\theta$-functions it is apparent that Landau
damping only occurs for  color fields with space-like wave vectors.
This is also true for an Abelian plasma~\cite{LL}.

Evaluation of the real part of the polarization tensor (\ref{realpo})
yields
\begin{mathletters}
\begin{eqnarray}
{\rm Re} \, \Pi^{00}_{ab} (\omega, {\bf k}) & = &\delta_{ab} \, \Pi_{l}
(\omega, {\bf k}) \ ,\\[2mm]
{\rm Re} \, \Pi^{0i}_{ab} (\omega, {\bf k}) & = & \delta_{ab} \,
\omega \, \frac{k^i}{|{\bf k}|^2}\, \Pi_{l} (\omega, {\bf k}) \ , \\
{\rm Re} \, \Pi^{ij}_{ab} (\omega, {\bf k}) & = & \delta_{ab} \left[
 \left ( \delta^{ij}- \frac{k^i k^j}{|{\bf k}|^2} \right) \Pi_{t} (\omega,
{\bf k})+ \frac{k^i k^j} {|{\bf k}|^2} \, \frac{\omega^2 }{|{\bf k}|^2} \,
\Pi_{l} (\omega, {\bf k}) \right] \ ,
\end{eqnarray}
\label{resultreal}
\end{mathletters}
$\!\!$where
\begin{mathletters}
\begin{eqnarray}
\Pi_{l} (\omega, {\bf k}) & = & m^2 _{D} \left( \frac{\omega}{2|{\bf
k}|} \,{\rm ln\,}\left|{\frac{\omega+|{\bf k}|}{\omega-|{\bf k}|}}\right|
-1  \right) \ , \\
 \Pi_{t} (\omega, {\bf k}) & = &- m^2 _{D} \, \frac{\omega^2}{2 |{\bf
k}|^2} \left[ 1 + \frac12 \left( \frac{|{\bf k}|}{\omega} -
\frac{\omega}{|{\bf k}|} \right) \,  {\rm ln\,} \left|{\frac{\omega+
|{\bf k}|}{\omega-|{\bf k}|}}\right| \, \right] \ .
\end{eqnarray}
\label{pipi}
\end{mathletters}
$\!\!$Equations (\ref{resultreal})--(\ref{pipi}) characterize Debye
screening, as well as longitudinal and transverse plasma waves.

Our results for the HTLs of the polarization tensor agree with those
obtained in the high temperature limit using quantum field theoretic
techniques\footnote{The connection between the retarded
polarization tensor computed here and the time-ordered polarization
tensor that is commonly used in quantum field theory  has been
studied in \cite{JN}.}~\cite{Weldon,BP1,FT,JN,BI,BI2}. We emphasize
that the above results are gauge-independent, and  obey the Ward
identity
\begin{equation}
k_{\mu} \Pi^{\mu\nu}_{ab} = 0 \ ,
\end{equation}
as should be expected from the gauge invariance of our formalism.

Previous applications of classical transport theory to QCD have
utilized an Abelian-dominance approximation to compute the
polarization tensor~\cite{EH}. It is noteworthy that there  one
recovers the same values of the polarization tensor that we found
here. The reason for this agreement is that the leading-order
contribution to the color current is made linear in the gauge field by
the  plane-wave {\it Ansatz}~\cite{BI2}, exactly as happens in the
Abelian-dominance approximation. However, the Abelian-dominance
approximation cannot  give a proper account of the {\it whole} set of
HTLs, such as thermal corrections to $n$-point functions, $n \ge 3$.

\section{CONCLUSIONS}
\label{sec5}

In this paper, we have shown how classical transport theory can be
used to derive the hard thermal loops of QCD. This formalism, we
believe, is more direct and transparent than previous approaches
based on perturbative quantum field theory. Indeed, hard thermal
loops represent UV-finite thermal corrections to propagators and
vertices. They arise from thermal scattering within a hot assembly of
particles, and one may reasonably expect them to be describable in
terms of classical physics.

The fact that we are modelling the high-temperature, deconfined,
phase of QCD allows us to treat color classically, and enables the
colored constituents of the plasma to be identified as quarks and
gluons. Employing classical transport theory to study colored
particles requires incorporating the color degrees of freedom into
phase-space. A consistent measure must be defined over the new
color coordinates. Furthermore, conservation of the group Casimirs
under the dynamical evolution must be ensured. One means to
accomplish these goals is to include delta-function constraints into
the phase-space volume element. We have formally justified this {\it
ad hoc} procedure by relating the dependent color degrees of
freedom to a set of independent Darboux canonical variables, and by
proving that the corresponding volume elements are equivalent.

A system of non-Abelian Vlasov equations describes transport
phenomena in the QCD plasma. This system governs the evolution of
both the single-particle phase-space distribution functions and the
mean color fields. It would be a formidable task to solve the
transport equation in the most general case, hence suitable
approximations must be made.

First, we specialize to a collisionless plasma, in which there is no
direct scattering between particles. This situation is not devoid of
interest since collective mean-field effects can, and indeed do, arise.

Second, we employ a perturbative approximation scheme. We
assume that the plasma is near equilibrium and expand the
phase-space distribution function in powers of $g$, the gauge
coupling constant. At the high temperature which must prevail for
the formation of a quark--gluon plasma, the thermal energies of the
particles are sufficiently large that the effects of the interactions
with the gauge fields are comparatively small, and we expect
perturbation theory to be valid.

Taking the high temperature limit constitutes our third
approximation. In this limit, the masses of the particles can be
neglected. Furthermore, the plasma is in a highly degenerate state, so
that the equilibrium distribution functions are determined by the
spin--statistics theorem.

We demand that gauge invariance be preserved by our perturbative
expansion. In its lowest, non-trivial, order this expansion leads to the
generating functional of HTLs. That gauge invariance is the
appropriate guiding principle in uncovering hard thermal loops is not
surprising. To apply this principle to the quantum field theoretic
calculation of HTLs involves gauge fixing, ghosts, and resummation
of classes of Feynman diagrams. In contrast, the route that one
has to follow in order to adhere to the gauge principle is
straightforward within classical physics. Therefore, the gauge
invariance property of hard thermal loops is self-evident in our
formalism.

\vspace{15mm}
{\bf Acknowledgements:} We thank Professor R. Jackiw for constant
encouragement and for many useful and enjoyable discussions.

\vspace{10mm}
\appendix
\section{Microscopic description}
\label{appA}

In a microscopic description, the particle's trajectory in phase-space
is known exactly. With this knowledge, we can construct a
distribution function $f(x,p,Q)$ (without loss of generality, we shall
consider only one particle):
\begin{equation}
f(x,p,Q) = \int {d\tau\over m}\ \delta^{(4)}\Bigl( x-x(\tau)\Bigr)\
\delta^{(4)}\Bigl( p-p(\tau)\Bigr)\ \delta^{(N^2-1)}\Bigl( Q-
Q(\tau)\Bigr)\ \ ,
\label{eq:microf}
\end{equation}
where $x(\tau),\ p(\tau)$ and $Q(\tau)$ obey the Wong equations
(\ref{wongeq}), {\it i.e.} they naturally fulfill the mass-shell and
Casimir constraints. For convenience, those constraints are here
subsumed into the distribution function instead of being contained
in the phase-space volume element. Had we not done this,
(\ref{eq:microf}) would have to be written in terms of both a
3-dimensional $\delta$-function in momentum space,
and an $N(N-1)$-dimensional $\delta$-function in color space.

We now prove that the expression (\ref{eq:microf}) for $f$ satisfies
the collisionless Boltzmann equation (\ref{boltzmann}). The first term
in (\ref{boltzmann}) can be rewritten, by using the properties of the
$\delta$-function, as:
\begin{equation}
p^\mu{\partial\over\partial{x^\mu}} f =-\int {d\tau\over m}\,
p^\mu(\tau)\left[{\partial\over\partial x^\mu (\tau)}\,
\delta^{(4)}\Bigl( x-x(\tau)\Bigr)\right]\,\delta^{(4)}\Bigl( p-
p(\tau)\Bigr)\, \delta^{(N^2-1)}\Bigl( Q-Q(\tau)\Bigr)\ .
\end{equation}
{}From this, after using the Wong equation for the variable
$x^\mu(\tau)$ and applying the chain rule, we get:
\begin{equation}
p^{\mu}{{\partial}\over{\partial x^{\mu}}} f = -\int d \tau\,
\left[{d \over d\tau}\ \delta^{(4)}\Bigl( x-x(\tau)\Bigr)\right]\,
\delta^{(4)}\Bigl( p-p(\tau)\Bigr)\, \delta^{(N^2-1)}
\Bigl( Q-Q(\tau)\Bigr)\ .
\label{eq:cond1}
\end{equation}
Similar arguments yield, for the second term in the Boltzmann
equation:
\begin{equation}
-g\,p^{\mu}Q_{a}F^{a}_{\mu\nu}{{\partial}\over{\partial p_{\nu}}} f
= -\int d \tau\,\delta^{(4)}\Bigl(x-x(\tau)\Bigr)
\left[{d \over d\tau}\delta^{(4)}\Bigl(p-p(\tau)\Bigr)\right]
\delta^{(N^2-1)}\Bigl(Q-Q(\tau)\Bigr)\ ,
\label{eq:cond2}
\end{equation}
and for the third term:
\begin{equation}
-g\,p^{\mu} f_{abc}A^{b}_{\mu} Q^{c}{{\partial}\over{\partial Q_{a}}}
f= -\int d \tau\,\delta^{(4)}\Bigl( x-x(\tau)\Bigr)\ \delta^{(4)}
\Bigl( p-p(\tau)\Bigr)\ \left[{d \over d\tau}\ \delta^{(N^2-1)}
\Bigl( Q-Q(\tau)\Bigr)\right]\ .
\label{eq:cond3}
\end{equation}
Adding together the equations (\ref{eq:cond1}), (\ref{eq:cond2}) and
(\ref{eq:cond3}), we observe that the left hand side -- the
$\tau$-integral of a total $\tau$-derivative -- vanishes, thereby
yielding the collisionless Boltzmann equation (\ref{boltzmann}).

\section{CONSERVATION OF THE COLOR CURRENT}
\label{appB}

Let us verify that the color current (\ref{cr5}) is covariantly
conserved. Using the collisionless transport equation, one can
compute
\begin{eqnarray}
\partial_{\mu}\, j^{\mu\,a} (x) & = & g \int dP \,dQ\, p^{\mu}  Q^a
\partial_{\mu} f(x,p,Q) \nonumber\\
& = & g^2 \int dP\,dQ\, p^{\mu} Q^a   \left(
Q_b F^{b} _{\mu\nu}(x) \frac{\partial}{\partial p_{\nu}} +
f^{dbc} A_{\mu\,b} (x) Q_{c} \frac{\partial}{\partial Q^{d}}
\right) f(x,p,Q) \ ,
\end{eqnarray}
where the color measure is $dQ = d^{(N^2-1)} Q\,\, C(Q)$, and $C(Q)$
specifies the color constraints in phase-space. Integrating by parts
and discarding surface terms, one gets
\begin{eqnarray}
\partial_{\mu}\, j^{\mu\,a} (x) &=& -g^2 \int dP \Biggl[ \int dQ \Bigl(
 \, Q^a Q_b g^{\mu}_{\nu} F^{b}_{\mu\nu} (x)
+ p^{\mu}  \delta_{cd} f^{dbc} A_{\mu\,b} (x)
Q^{a}  + p^{\mu}  f^{dbc} A_{\mu\,b} (x) \delta^a_d Q_c \Bigr)
\nonumber\\
&&+  \int d^{(N^2-1)} Q \  p^{\mu} A_{\mu\,b} (x) f^{dbc} Q_{c}
 \frac{\partial}{\partial Q^{d}} C(Q) \Biggr] f(x,p,Q) \ .
\end{eqnarray}
Among the four terms in the right side, only the third one survives.
The first two terms cancel due to antisymmetry of $F^a _{\mu\nu}$
and $f^{dbc}$, respectively. The last term also cancels, since the
constraints $C(Q)$ are gauge-invariant, {\it i.e.},
\begin{equation}
0 = \delta Q^a \, \frac{ \delta C(Q)} {\delta Q^a} = -g f^{abc} \epsilon_b
(x) Q_c \, \frac{\partial C(Q)}{\partial Q^a} \ ,
\end{equation}
where $\delta Q^a$ denotes an infinitesimal gauge transformation
with arbitrary parameter $\epsilon_b (x)$. (For $SU(3)$ this last
property can be explicitly checked by using the Jacobi-like identity
$f_{abc}d_{dec} + f_{adc}d_{ebc} + f_{aec}d_{bdc}=0$.)
Finally, one obtains the expression for the covariant conservation of
the color current: $\partial_{\mu} \,j^{\mu\,a} (x) + g f^{abc}
A_{\mu\,b}(x) j^{\mu}_c (x) =0$.

\end{document}